\documentclass[aps,preprint,tightenlines,showpacs]{revtex4}
\usepackage{graphicx}
\usepackage{amsmath}
\usepackage{amssymb}

\begin{document}

\title{Characterization of shell filling of interacting polarons in a quantum dot
through their optical absorption }%
\author{S. N. Klimin$^{*}$, V. M. Fomin$^{*,**}$, F. Brosens, and J. T.
Devreese$^{\sharp,\,\flat}$}%
\affiliation{Theoretische Fysica van de Vaste Stoffen (TFVS), Universiteit
Antwerpen (UIA), B-2610 Antwerpen, Belgium}%

\begin{abstract}
The method for calculating the ground-state energy and the optical
conductivity spectra is developed for a system of a finite number of
interacting arbitrary-coupling polarons in a spherical quantum dot with a
parabolic confinement potential. The path-integral formalism for identical
particles is used in order to take into account the fermion statistics.
Using a generalization of the Jensen-Feynman variational principle, the
ground-state energy of a confined $N$-polaron system is analyzed as a
function of $N$ and of the electron-phonon coupling strength. The
calculated optical conductivity spectra of the $N$-polaron system in a
quantum dot manifest features related to ground-state transitions between
states with different total spin.
\end{abstract}

\date{\today}
\maketitle


\section{Introduction}

Many-electron states in quantum dots are intensely investigated by various
theoretical methods (see, e. g., Refs.
\cite{PRB61-1971,P2000-2,PRL82-3320,PRB62-8108}). These works do not take
into account the electron-phonon interaction, whereas it can contribute
significantly to both the equilibrium and the non-equilibrium properties
of quantum dots. In particular, the electron-phonon interaction plays a
key role in the optical spectra of some quantum dots (see, e.~g., Ref.
\cite{PRB57-2415} and references therein). To the best of our knowledge,
the many-polaron optical absorption in quantum dots has not yet been
widely studied. In Refs. \cite{SSC114-305,SPRB}, the ground state and the
optical response of a fixed number of identical interacting polarons are
analyzed using the variational path-integral method for identical
particles \cite{PRE96,PRE97,Note}. As far as we investigate a system with
a fixed number of identical particles, it should be described using the
\emph{canonical} ensemble \cite{Note}. As distinct from Ref.
\cite{SSC114-305}, where only closed-shell systems are considered, the
present approach is developed for both closed-shell and open-shell
systems.

\section{Theoretical approach}

We consider a system of $N$ electrons with Coulomb repulsion, which
interact with the lattice vibrations. We assume a spherical quantum dot
with a parabolic confinement potential characterized by the frequency
$\Omega_{0}$. The partition function of the system is given by a path
integral over electron and phonon coordinates (for details, see Ref.
\cite{SPRB}). After elimination of the phonon paths, the partition
function of the electron-phonon system factorizes into a product of a
free-phonon partition function with a partition function $Z_{p}\left(
\left\{ N_{\sigma}\right\} ,\beta\right)  $ [where $\beta\equiv1/\left(
k_{B}T\right)  $] of interacting
polarons, which is a path integral over the electron coordinates only:%
\begin{align}
Z_{p}\left(  \left\{  N_{\sigma}\right\}  ,\beta\right)   &  =\sum
_{P}\frac{\left(  -1\right)  ^{\mathbf{\xi}_{P}}}{N_{1/2}!N_{-1/2}%
!}\nonumber\\
&  \times\int d\mathbf{\bar{x}}\int_{\mathbf{\bar{x}}}^{P\mathbf{\bar{x}}%
}D\mathbf{\bar{x}}\left(  \tau\right)  e^{-S_{p}\left[  \mathbf{\bar{x}%
}\left(  \tau\right)  \right]  },\label{Zp}%
\end{align}
where $S_{p}\left[  \mathbf{\bar{x}}\left(  \tau\right)  \right]  $
contains the so-called influence phase of the phonons. It describes the
phonon-induced retarded interaction between the electrons, including the
retarded self-interaction of each electron with itself. The free energy of
a system of interacting polarons $F_{p}\left( \left\{  N_{\sigma}\right\}
,\beta\right)  $ is related to the partition function (\ref{Zp}) by the
expression
\begin{equation}
F_{p}\left(  \left\{  N_{\sigma}\right\}  ,\beta\right)
=-\frac{1}{\beta}\ln
Z_{p}\left(  \left\{  N_{\sigma}\right\}  ,\beta\right)  .\label{Fp}%
\end{equation}

At present no method is known to calculate the non-gaussian path integral
(\ref{Zp}) analytically. For \emph{distinguishable} particles, the
Jensen-Feynman variational principle \cite{Feynman} provides a powerful
approximation technique. It yields a lower bound to the partition
function, and hence an upper bound to the free energy.

It is a non-trivial problem, how a variational principle for the free
energy should be formulated for a system of \emph{identical}
\emph{particles}. However, it was shown \cite{PRE96} that the
path-integral approach to the many-body problem for a fixed number of $N$
identical particles could be formulated in terms of a Feynman-Kac
functional on a state space for $N$ indistinguishable particles, by
imposing an ordering on the configuration space, and the introduction of a
set of boundary conditions at the boundaries of this state space. The path
integral (with the imaginary-time variable) for identical particles was
shown to be \emph{positive} within this state space. Thus a many-body
extension of the Jensen-Feynman inequality was found, which could be used
for interacting identical particles (Ref. \cite{PRE96}, p. 4476). A more
detailed analysis of this variational principle for both local and
retarded interactions can be found in Ref. \cite{Note}. It is required
that the potentials are symmetric with respect to all permutations of the
particle positions, and that both the exact propagator and the model
propagator are antisymmetric (for fermions) with respect to permutations
of any two electrons at any time. This means that both propagators have to
be defined on the same configuration space. Under these requirements, the
variational inequality for identical particles has the same form as that
for the Jensen-Feynman variational principle:
\begin{equation}
F_{p}\leqslant F_{0}+\frac{1}{\beta}\left\langle S_{p}-S_{0}\right\rangle
_{S_{0}},\label{JF}%
\end{equation}
where $S_{0}$ is a model action with the corresponding free energy
$F_{0}$.
The angular brackets mean a weighted ave-rage over the paths%
\begin{align}
\left\langle \left(  \bullet\right)  \right\rangle _{S_{0}} &  =\frac{1}%
{Z_{p}\left(  \left\{  N_{\sigma}\right\}  ,\beta\right)  }\sum_{P}%
\frac{\left(  -1\right)  ^{\mathbf{\xi}_{P}}}{N_{1/2}!N_{-1/2}!}\nonumber\\
&  \times\int d\mathbf{\bar{x}}\int_{\mathbf{\bar{x}}}^{P\mathbf{\bar{x}}%
}D\mathbf{\bar{x}}\left(  \tau\right)  \left(  \bullet\right)  e^{-S_{0}%
\left[  \mathbf{\bar{x}}\left(  \tau\right)  \right]  }.\label{Aver}%
\end{align}
In the present work, we have chosen a model system consisting of $N$
electrons with coordinates $\mathbf{\bar{x}\equiv}\left\{
\mathbf{x}_{j,\sigma }\right\}  $ and $N_{f}$ fictitious particles with
coordinates $\mathbf{\bar {y}\equiv}\left\{  \mathbf{y}_{j}\right\}  $ in
a harmonic confinement potential with elastic interparticle interactions
as studied in Ref. \cite{SSC114-305}. The model confinement frequencies
for an electron and for a fictitious particle, the force constants and the
mass of a fictitious particle are the variational parameters.

In order to investigate the optical properties of the many-polaron system,
we extend in Ref. \cite{SPRB} the memory-function formalism developed in
Ref. \cite{DSG72} to the case of interacting polarons in a quantum dot.
Within this technique, the optical conductivity for a system of
interacting polarons in a parabolic confinement potential is given in
terms of the memory function
$\chi\left(  \omega\right)  $,%
\begin{equation}
\operatorname{Re}\sigma\left(  \omega\right)
=-\frac{e^{2}}{m}\frac{\omega \operatorname{Im}\chi\left(  \omega\right)
}{\left[  \omega^{2}-\Omega _{0}^{2}-\operatorname{Re}\chi\left(
\omega\right)  \right]  ^{2}+\left[
\operatorname{Im}\chi\left(  \omega\right)  \right]  ^{2}},\label{Kw}%
\end{equation}
where $m$ is the electron band mass, and $\chi\left(  \omega\right)  $ is
\begin{align}
\chi\left(  \omega\right)   &  =\sum_{\mathbf{q}}\frac{2\left|  V_{\mathbf{q}%
}\right|  ^{2}q^{2}}{3N\hbar m}\int\limits_{0}^{\infty}\left(  e^{i\omega
t}-1\right)  \nonumber\\
&  \times\operatorname{Im}\left[  T_{\omega_{\mathrm{LO}}}^{\ast}\left(
t\right)  \left\langle \rho_{\mathbf{q}}\left(  t\right)  \rho_{-\mathbf{q}%
}\left(  0\right)  \right\rangle _{M}\right]  \,dt.\label{memfun}%
\end{align}
Here, $V_{\mathbf{q}}$ is the amplitude of the electron-phonon interaction
\cite{DSG72}, $T_{\omega}\left(  t\right)  =\cos\left[  \omega\left(
t-i\hbar\beta/2\right)  \right]  /\sinh\left(  \beta\hbar\omega/2\right) $
is the phonon Green's function, and $\left\langle \rho_{\mathbf{q}}\left(
t\right)  \rho_{-\mathbf{q}}\left(  0\right) \right\rangle _{M}$ is the
density-density correlation function calculated using the model Lagrangian
of electrons harmonically interacting with fictitious particles. It is
noteworthy that the optical conductivity (\ref{Kw}) differs from that for
a translationally invariant polaron system both by the explicit form of
$\chi\left(  \omega\right)  $ and by the presence of the term
$\Omega_{0}^{2}$ in the denominator.

\section{Results and discussion}

\begin{figure}
[ptbh]
\begin{center}
\includegraphics[
width=4in
]%
{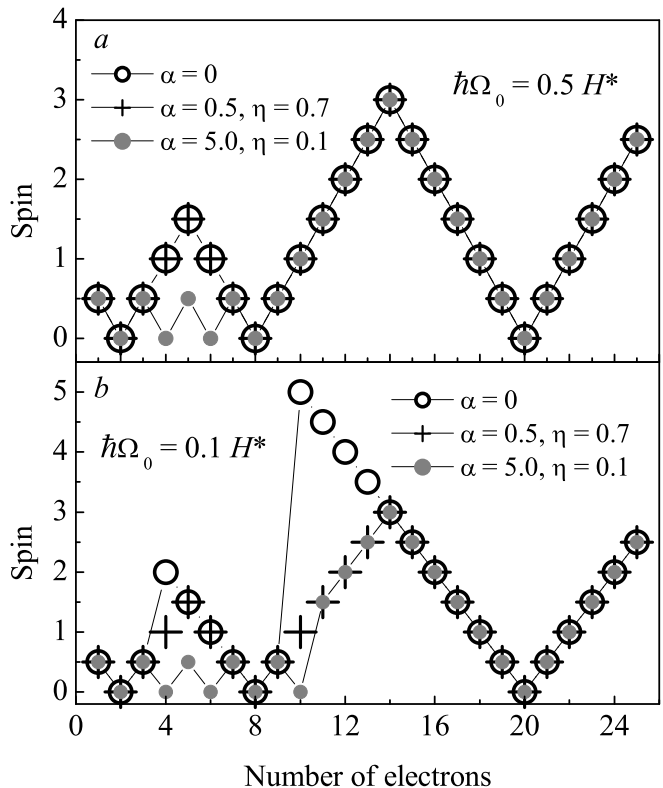}%
\caption{Total spin of the system of interacting polarons in a quantum dot
as a function of the number of electrons for the confinement energy $\hbar
\Omega_{0}=0.5H^{\ast}$ (\emph{a}) and for $\hbar\Omega_{0}=0.1H^{\ast}$
(\emph{b}). The confinement energy is measured in effective Hartrees
$H^{\ast }=\left[  m/\left(  m_{0}\varepsilon_{\infty}^{2}\right)  \right]
$ Hartree,
where $m_{0}$ is the electron mass in the vacuum.}%
\label{Spin}%
\end{center}
\end{figure}

Further on, we present numerical results for the many-polaron system in a
quantum dot at $T=0$. The total spin of interacting polarons in its ground
state, as well as the total spin of a confined few-electron system,
depends on the number of electrons. The analysis of this dependence
reveals a shell structure of quantum dots. In Fig. \ref{Spin}, the total
spin $S$ of a finite number of interacting polarons in a spherical quantum
dot for different values of the confinement energy $\hbar\Omega_{0}$, of
the electron-phonon coupling constant $\alpha$ and of the ratio of the
high-frequency and the static dielectric constants
$\eta\equiv\varepsilon_{\infty}/\varepsilon_{0}$ is plotted as a function
of the number of electrons in a quantum dot. As distinct from few-electron
systems without the electron-phonon interaction, three types of spin
polarization are possible for the ground state, which can be
distinguished from each other using, e. g., capacity measurements.%

(i) Except the strong-coupling regime and the low-density case, for
closed-shell systems $S=0$, while for open-shell systems $S$ takes a
maximal value for a given shell filling, in accordance with the Hund's
rule.

(ii) When weakening confinement for a fixed number of electrons, the
electron density lowers. Hence, at sufficiently small values of
$\Omega_{0}$, a spin-polarized state for a system of interacting polarons
in a quantum dot becomes energetically more favorable than a state
satisfying the Hund's rule.

(iii) In the strong-coupling regime ( $\alpha\gg1$ and $\eta\ll1$), the
total spin of an open-shell system for the ground state can take a minimal
possible value. This trend to minimize the total spin is a consequence of
the electron-phonon interaction, presumably due to the fact that the
phonon-mediated electron-electron attraction overcomes the Coulomb
repulsion.

The optimal values of the variational parameters are used as input for the
calculation of the optical-conductivity spectrum of the system (\ref{Kw}).
The ground-state transitions between the aforesaid states with different
values of the total spin, which occur when changing the number of
electrons, are manifested in the optical absorption spectra. In
particular, the shell structure for a system of interacting polarons in a
quantum dot is clearly pronounced when
analyzing the first frequency moment of the optical conductivity,%
\begin{equation}
\left\langle \omega\right\rangle =\frac{\int_{0}^{\infty}\omega
\operatorname{Re}\sigma\left(  \omega\right)  d\omega}{\int_{0}^{\infty
}\operatorname{Re}\sigma\left(  \omega\right)  d\omega}.\label{Fmom}%
\end{equation}

\begin{figure}
[ptbh]
\begin{center}
\includegraphics[
width=4in
]%
{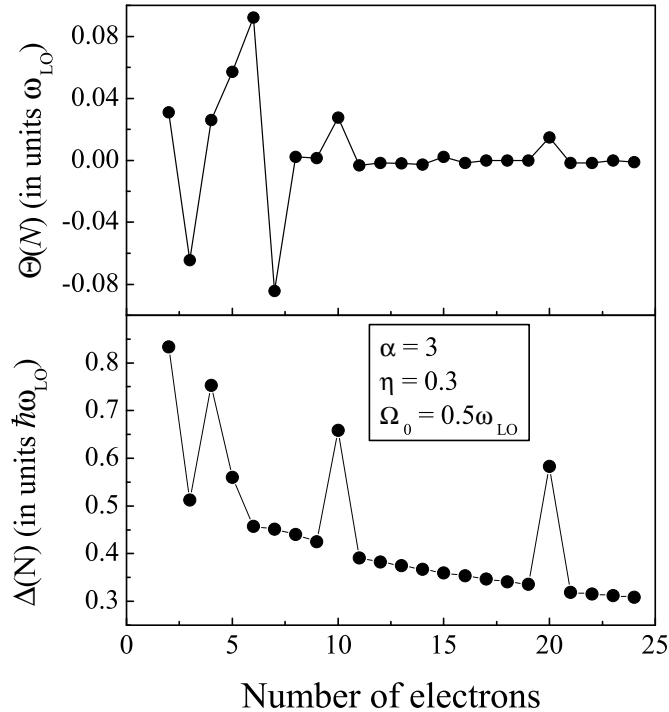}%
\caption{The function $\Theta\left(  N\right)  $ (\emph{a}) and the
addition energy $\Delta\left(  N\right)  $ (\emph{b}) for systems of
interacting polarons in a quantum dot with $\alpha=3,$ $\eta=0.3$ and
$\Omega
_{0}=0.5\omega_{\mathrm{LO}}$ ($\hbar\Omega_{0}\approx0.01361$ $H^{\ast}$).}%
\label{Moments}%
\end{center}
\end{figure}

Figure \ref{Moments} represents the function%
\begin{equation}
\Theta\left(  N\right)  \equiv\left.  \left\langle \omega\right\rangle
\right|  _{N+1}-2\left.  \left\langle \omega\right\rangle \right|
_{N}+\left.  \left\langle \omega\right\rangle \right|  _{N-1}\label{Theta}%
\end{equation}
along with the addition energy $\Delta\left(  N\right)  $ needed to put an
extra electron into the quantum dot with $N$ electrons,
\begin{equation}
\Delta\left(  N\right)  =E^{0}\left(  N+1\right)  -2E^{0}\left(  N\right)
+E^{0}\left(  N-1\right)  ,\label{Add}%
\end{equation}
where $E^{0}\left(  N\right)  $ is the ground-state energy.%

Both functions $\Theta\left(  N\right)  $ and $\Delta\left(  N\right)  $
exhibit features at $N=4$ and at $N=7$, where the aforesaid transition
occurs. Namely, at $N=4$, $\Theta\left(  N\right)  $ and $\Delta\left(
N\right)  $ have minima, whereas at $N=7$, $\Theta\left(  N\right)  $ has
a minimum, and $\Delta\left(  N\right)  $ has a kink. Pronounced peaks
appear in these functions at the ``magic numbers'' $N=10$ and $N=20$,
which correspond to closed-shell systems. The transition between the
spin-polarized ground state and the ground state obeying the Hund's rule
thus should be observable using optical measurements.

\section{Conclusions}

The path-integral treatment of the quantum statistics of indistinguishable
particles allowed us to find an upper bound to the ground-state energy of
a fixed number of polarons in a parabolic confinement potential for
arbitrary electron-phonon coupling strength.

The transitions between states with different values of the total spin
manifest themselves through discontinuous changes in the optical
conductivity spectra and of the addition energy as a function of the
number of electrons. Thus, the analysis of the optical-conductivity
spectra provides a tool for examining the shell structure of a system of
interacting polarons through experimental measurements of the optical
absorption in quantum dots.

This work has been supported by the GOA BOF UA 2000, I.U.A.P., F.W.O.-V.
projects G.0274.01, G.0435.03, the W.O.G. WO.025.99 (Belgium) and the
European Commission GROWTH Programme, NANOMAT project, contract No.
G5RD-CT-2001-00545.

\end{document}